\documentclass[preprintnumbers,nofootinbib,preprint,12pt,showpacs,amsmath,amssymb]{revtex4}
\usepackage{bm}
\usepackage{graphicx}
\begin{document}

{~}
\vspace{3cm}
\title{\Large Topology Change of Coalescing Black Holes \\ on Eguchi-Hanson Space
\vspace{1cm}}
\author{
Hideki Ishihara\footnote{E-mail:ishihara@sci.osaka-cu.ac.jp}, 
Masashi Kimura\footnote{E-mail:mkimura@sci.osaka-cu.ac.jp}, 
and
Shinya Tomizawa\footnote{E-mail:tomizawa@sci.osaka-cu.ac.jp}
}

\affiliation{
Department of Mathematics and Physics,
Graduate School of Science, Osaka City University,
3-3-138 Sugimoto, Sumiyoshi, Osaka 558-8585, Japan
\vspace{2cm}}

\begin{abstract}
We construct multi-black hole solutions in the five-dimensional Einstein-Maxwell theory with a positive cosmological constant on the Eguchi-Hanson space, which is an asymptotically locally 
Euclidean space.
The solutions describe the physical process such that two black holes 
with the topology of ${\rm S}^3$ coalesce into a single black hole with the topology of the lens space 
$L(2;1)={\rm S}^3/{\mathbb Z}_2$. We discuss how the area of the single black hole 
after the coalescence depends on the topology of the horizon.
\end{abstract}

\preprint{OCU-PHYS 255}
\preprint{AP-GR 37}

\date{\today}
\maketitle

In recent years, the studies on black holes in higher dimensions
have attracted much attention in the context of string theory
and the brane world scenario.
In fact, it is expected that higher dimensional black holes
would be produced in a future linear collider~\cite{BHinCollider}.
Such physical phenomena are expected not only to give us
a piece of evidence for the existence of extra dimensions
but also to help us to draw some information toward quantum gravity.

Some of studies on higher-dimensional black holes
show that they have much more complicated and richer structure
than four-dimensional ones.
For an example, in asymptotically flat spacetimes,  the topology of the event horizon
in higher dimensions cannot be uniquely determined~\cite{Cai,HelfgottGalloway}
in contrast to four-dimensional ones, which is restricted only to ${\rm S}^2$.
In five dimensions, the possible horizon topology
is either ${\rm S}^3$ or ${\rm S}^1\times {\rm S}^2$~\cite{Cai}.
In fact, the black hole solutions with these topologies were found as vacuum solutions in five-dimensional Einstein equations~\cite{EmparanMyers}.
In dimensions higher than five, black holes will have more complicated structure~\cite{HelfgottGalloway}. 

Black hole spacetime which has asymptotically Euclidean time slices, AE black holes in short, would be a good idealization 
in the situation such that we can ignore the tension of the brane 
and the curvature radius of the bulk, {or the size of extra dimensions. 
However, from more realistic view point, we need not impose the asymptotic Euclidean condition 
toward the extra dimensions. 
In fact, higher dimensional black holes admit a variety of asymptotic structures: Kaluza-Klein black hole solutions~\cite{IM,IKMT} have the spatial infinity 
with compact extra dimensions; Black hole solutions on the Eguchi-Hanson space~\cite{IKMT2} have the spatial infinity of topologically various lens spaces $L(2n;1)={\rm S}^3/{\mathbb Z}_{2n}$ ($n$:natural number). The latter black hole spacetimes are asymptotically locally Euclidean, {i.e.}, ALE black holes.
In spacetimes with such asymptotic structures, furthermore, 
black holes have the different structures from the black hole 
with the asymptotically Euclidean structure. 
For instance, the Kaluza-Klein black holes~\cite{IM,IKMT} 
and the black holes on the Eguchi-Hanson space~\cite{IKMT2} can have the horizon of lens spaces in addition to ${\rm S}^3$.

In order to verify the existence of the extra dimensions, we need to know 
the properties of higher dimensional spacetimes and observe the physical 
phenomena caused by the effect of the extra dimensions. The phenomena would depend on asymptotic structure of the spacetime including the extra dimensions. As an example, we will investigate how to coalescence of five-dimensional two black holes depends on the asymptotic structure of spacetime. It is important to get information of the extra dimensions from higher dimensional black holes in our effective spacetime.

We compare five-dimensional static and vacuum AE
black holes (Schwarzschild black holes) with 
ALE black holes 
which has the spatial infinity with the topology 
of a lens space $L(2;1)={\rm S}^3/{\mathbb Z}_2$.
The metric of an ALE black hole we consider is given by 
\begin{eqnarray}
ds^2 = -\bigg(1-\frac{r_g^2}{r^2}\bigg)dt^2 
+ \bigg(1-\frac{r_g^2}{r^2}\bigg)^{-1}dr^2 +r^2 d\Omega_{{\rm S}^3/{\mathbb Z}_2}^2,\label{eq:1}
\end{eqnarray}
where $r_g$ is a constant, and the metric of the ${\rm S}^3$ in the Schwarzschild spacetime is replaced by the metric 
of the lens space, which has the locally same geometry as ${\rm S}^3$.

The ADM mass of the ALE black hole~(\ref{eq:1}) is half of 
the mass of Schwarzschild black hole with the same $r_g$.
It means that the area of horizon of the
ALE black hole is $\sqrt{2}$ times that of the Schwarzschild black hole
with the same ADM mass.
This fact would cause the difference of the
coalescence process of higher dimensional
black holes with non trivial asymptotic structure.

To observe this, 
we compare the ALE multi-black hole solutions constructed on 
the Eguchi-Hanson space with the five-dimensional 
version of Majumdar-Papapetrou solutions~\cite{M-PMyers}, which has AE structure.
We can prepare a pair of black holes with the same 
mass and area of ${\rm S}^3$-horizons in both ALE and AE solutions.
If the black holes with ${\rm S}^3$ topology coalesce into a single black holes,
it would be natural that the resultant black hole has ${\rm S}^3$-horizon 
in the AE case while $L(2;1)$-horizon in the ALE case,
because there are closed surfaces with ${\rm S}^3$ topology 
surrounding two black holes in the AE case, 
while $L(2;1)$ topology in the ALE case.
If we assume the total mass of black holes is conserved through 
the whole process, the area of the final 
black hole in the ALE case is larger than that in the AE case.

Kastor and Traschen~\cite{KT} constructed multi-black hole solutions 
with a positive cosmological constant,  
and London~\cite{London} extend it to five-dimensional solutions.
If we consider a contracting phase derived by the cosmological constant, 
the solution can describe a 
coalescence of black holes in an asymptotically de Sitter spacetime~\cite{BHKT,IN}.
Analogously, we investigate the coalescence of the ALE black holes by using new  multi-black hole solutions on the Eguchi-Hanson 
space with a positive cosmological constant which is asymptotically locally de Sitter.

We consider the five-dimensional Einstein-Maxwell system with a positive cosmological constant described by the action,
\begin{equation}
S=\frac{1}{16\pi G_5}\int dx^5 \sqrt{-g} (R +\Lambda-F_{\mu\nu}F^{\mu \nu} ),
\end{equation}
where $R$ is the five dimensional scalar curvature, $\bm{F}=d \bm{A}$ 
is the five-dimensional Maxwell field strength tensor, 
$\Lambda$ is the positive cosmological constant and $G_5$ is the five-dimensional Newton constant. In the previous work~\cite{IKMT2}, we gave a metric of a pair of maximally charged black holes on the Eguchi-Hanson space as a solution in the five-dimensional Einstein-Maxwell theory without a positive cosmological constant.
The Einstein equation with a positive cosmological constant and the Maxwell equation admit a new solution whose metric and gauge potential one-form are
\begin{eqnarray}
ds^2&=&-H^{-2}d\tau^2+Hds_{\rm EH}^2,
\\
\bm{A}&=&\pm\frac{\sqrt{3}}{2}H^{-1}d\tau,
\end{eqnarray}
with
\begin{eqnarray}
ds_{\rm EH}^2&=&\biggl(1-\frac{a ^4}{r^4}\biggr)^{-1}dr^2
             +\frac{r^2}{4}\biggl[\biggl(1-\frac{a ^4}{r^4}\biggr) 
(d\tilde\psi+\cos\tilde\theta d\tilde\phi)^2+d\tilde\theta^2+\sin^2\tilde\theta d\tilde\phi^2\biggr],
\label{EguchiHanson}
\\
H
&=&
\lambda\tau+\frac{2m_1}{r^2-a ^2\cos\tilde{\theta}}
+\frac{2m_2}{r^2+a ^2\cos\tilde{\theta}},
\end{eqnarray}
where $a $ and $m_j\ (j=1,2)$ are positive constants,
$\lambda $ is a constant related to the cosmological constant by 
$\lambda ^2=4\Lambda/3$, $-\infty< \tau < \infty$, $a\le r < \infty$, $0\le \tilde\theta \le\pi$, $0\le \tilde\phi \le2\pi$ and $0\le \tilde\psi \le2\pi$.
Equation~(\ref{EguchiHanson}) is the metric form of the Eguchi-Hanson space~\cite{Eguchi}. As is seen later, this solution describes coalescing two black holes.

In order to obtain the physical interpretation about this solution, let us introduce the following coordinate~\cite{Prasad,IKMT2},
\begin{eqnarray}
R=a \sqrt{\frac{r^4}{a ^4}-\sin^2\tilde\theta},~
\tan\theta &=& \sqrt{1-\frac{a ^4}{r^4}}\tan\tilde\theta,
\\ 
\phi=\tilde\psi,~
\psi=2\tilde\phi,
\end{eqnarray}
where $0\le R < \infty$, $ 0\leq \theta\leq \pi,~
  0\leq \phi\leq 2\pi,~
  0\leq \psi\leq 4\pi $.
Then, the metric takes the form of 
\begin{eqnarray}
ds^2&=&-H^{-2}d\tau^2
    +H\bigg[V^{-1}dR^2
+V^{-1}R^2
(d\theta^2+\sin^2\theta d\phi^2)
+V
\bigg(\frac{a}{8}d\psi + \omega_{\phi}d\phi\bigg)^2
\bigg],\label{eq:EH}
\end{eqnarray}
with
\begin{eqnarray}
H &=&\lambda \tau+\frac{2m_1/a }{|{\bm R-\bm R_1}|}+\frac{2m_2/a }{|{\bm R-\bm R_2}|},\quad
\label{eq:harmo1}
\\
V^{-1}&=&\frac{a /8}{|{\bm R-\bm R_1}|}+\frac{a /8}{|{\bm R-\bm R_2}|},
\label{eq:harmo2}
\\
\omega_{\phi}&=&\frac{a}{8}
\bigg(
\frac{z - a}{|{\bm R-\bm R_1}|}
+
\frac{z + a}{|{\bm R-\bm R_2}|}
\bigg),
\label{eq:harmo3}
\end{eqnarray}
where $\bm{ R}=(x,y,z)$ is the position vector on the three-dimensional Euclid space and $\bm{R}_1=(0,0,a)$,
$\bm{R}_2=(0,0,-a)$. 
In order to focus our attention on the coalescence of two black holes, we consider only the contracting phase 
$\lambda=-\sqrt{4\Lambda/3}$ throughout below. Though $\tau$ runs the range $(-\infty,\infty)$, we investigate only the region $-\infty< \tau \le 0$.

For later convenience, we mention the global structure of the five-dimensional Reissner-Nordstr\"om-de Sitter solution with $m=\sqrt{3}|Q|/2$. This solution is static, spherically symmetric and has the horizons with the topology of ${\rm S}^3$. By the coordinate transformation into the cosmological coordinate, the metric is given by~\cite{London},
\begin{eqnarray}
ds^2 &=& -\bigg( \lambda  \tau + \frac{m}{r^2}\bigg)^{-2}d\tau^2 
+ \bigg( \lambda  \tau + \frac{m}{r^2}\bigg)
\bigg[
dr^2 
+ \frac{r^2}{4}d\Omega_{{\rm S}^2} + \frac{r^2}{4}(d\psi + \cos \theta d\phi )^2
\bigg],\label{eq:RNdS}
\end{eqnarray}
where each coordinate runs the range of
$-\infty<\tau < \infty,~
 0\le r< \infty,~
 0 \leq \theta \leq \pi,~
 0 \leq \phi \leq 2\pi,~
 0 \leq \psi \leq 4\pi,
$ $m$ is a constant, $\lambda$ is the constant related to the cosmological constant by $\lambda^2 = 4\Lambda/3$. The ingoing and outgoing expansions of the null geodesics orthogonal to three-dimensional surfaces, $\tau = {\rm const}$ and $r = {\rm const}$ are given by
\begin{eqnarray}
\theta_{\rm in} = \lambda  - \frac{2 x}{\sqrt{( x+m)^3}},\quad
\theta_{\rm out} = \lambda + \frac{2 x}{\sqrt{(x+m)^3}},
\end{eqnarray}
respectively, where $ x :=\lambda \tau r^2$.
There is a curvature singularity at $ x+m=0$. 
Horizons occur at $ x$ such that 
\begin{eqnarray}
\lambda ^2( x+m)^3-4 x^2=0.\label{eq:cubic}  
\end{eqnarray}
For $m < m_{\rm ext }\equiv 16/(27\lambda ^2)$, there are three horizons, 
which are three real roots $ x_{\rm in}[m]< x_{\rm BH}[m]< x_{\rm dS}[m]$, 
the inner and outer black hole horizons and the de Sitter horizon, respectively. 
We see that $\theta_{\rm out}=0$ at $ x= x_{\rm BH}[m]$ and $ x= x_{\rm dS}[m]$, and $\theta_{\rm in}=0$ at $ x= x_{\rm in}[m]$. When $m=m_{\rm ext}$, the outer black hole horizon and the de Sitter horizon coincides with each other ($x_{\rm BH}[m_{\rm ext}]=x_{\rm dS}[m_{\rm ext}]$). In the case of $m>m_{\rm ext}$ there is only a naked singularity.

Let us investigate the global structure of the solution~(\ref{eq:EH})-(\ref{eq:harmo3})
following the discussion  of Ref.\cite{BHKT}.
As mentioned below, in order to consider the coalescence of two black holes we must choose the parameters such that $m_1+m_2<8/(27\lambda ^2)$, $m_i>0$. Therefore, in this letter, we assume such range of these parameters. 

 First,
let us choose the origin on the three-dimensional Euclid space to be ${\bm R}={\bm R_i}\ (i=1,2)$ in Eqs.(\ref{eq:EH})-(\ref{eq:harmo3}). In the neighborhood of $R=0$ the metric becomes
\begin{eqnarray}
ds^2&\simeq &-\biggl(\lambda \tau+\frac{2m_i/a}{R}\biggr)^{-2}d\tau^2
\nonumber\\ &&
+\biggl(\lambda \tau+\frac{2m_i/a}{R}\biggr)
\frac{a}{8}
\biggl[
\frac{dR^2}{R}
+R d\Omega_{{\rm S}^2}^2
+R
(d\psi + \cos \theta d\phi)^2
\biggr],
\end{eqnarray} 
where the origin of $\tau$ is appropriately shifted by a constant. Using the coordinate $\tilde r^2=aR/2$, the metric can be written in the form,
\begin{eqnarray}
ds^2
&\simeq &
 -\biggl(\lambda \tau+\frac{m_i}{\tilde r^2}\biggr)^{-2}d\tau^2
+\biggl(\lambda \tau+\frac{m_i}{\tilde r^2}\biggr)
\biggl[d\tilde r^2+\frac{\tilde r^2}{4}d\Omega_{{\rm S}^2}^2+\frac{\tilde r^2}{4}(d\psi+\cos\theta d\phi)^2\biggr].\label{eq:near}
\end{eqnarray}
This is identical to the metric of the five-dimensional Reissner-Nordstr\"om-de Sitter solution~(\ref{eq:RNdS}) which has mass equal to $m_i$ which is written in the cosmological coordinate. 
If $m_i<16/(27\lambda ^2)$, which is automatically satisfied as long as we assume $m_1+m_2<8/(27\lambda ^2)$ and $m_i>0$, 
at early time $\tau \ll 0$, sufficiently small spheres with the topology of ${\rm S}^3$ centered at ${\bm R}={\bm R}_i$ are always outer trapped, since there are solutions for $\theta_{\rm out}=0$ at $\tilde r^2=x_{\rm BH}[m_1]/(\lambda\tau)$ and $\tilde r^2=x_{\rm BH}[m_2]/(\lambda\tau)$, which denote an approximate small sphere, respectively.

Next, we study the asymptotic behavior of the metric for large $R:=|{\bm R}|$,  where we assume that $R$ is much larger than the coordinate distance $|{\bm R_1}-{\bm R_2}|$ between the two masses $2m_1/a$ and $2m_2/a$. In this region, the metric behaves as
\begin{eqnarray}
ds^2& \simeq & -\biggl(\lambda \tau+\frac{2(m_1+m_2)/a}{R}\biggr)^{-2}d\tau^2
\nonumber\\ &&
+\biggl(\lambda \tau+\frac{2(m_1+m_2)/a}{R}\biggr)\frac{a}{4}\biggl[\frac{dR^2}{R}
+Rd\Omega_{{\rm S}^2}^2
+R
\bigg(\frac{d\psi}{2} + \cos \theta d\phi\bigg)^2
\biggr].
\end{eqnarray}
Here, let us introduce a new coordinate $r^2:=aR$, and then the metric takes the following form,
\begin{eqnarray}
ds^2& \simeq & -\biggl(\lambda \tau+\frac{2(m_1+m_2)}{r^2}\biggr)^{-2}d\tau^2\nonumber\\
    & &+\biggl(\lambda \tau+\frac{2(m_1+m_2)}{r^2}\biggr)\biggl[dr^2
+
\frac{r^2}{4}d\Omega_{{\rm S}^2}^2+\frac{r^2}{4}\biggl(\frac{d\psi}{2}+\cos\theta d\phi\biggr)^2\biggr].\label{eq:lens}
\end{eqnarray}
This resembles the metric of the five-dimensional Reissner-Nordstr\"om-de Sitter solution~(\ref{eq:RNdS}) with mass equal to $2(m_1+m_2)$. 
Like the five-dimensional Reissner-Nordstr\"om solution, if we assume $2(m_1+m_2)<16/(27\lambda ^2)$, 
at late time $\tau\to -0$, sufficiently large spheres becomes outer trapped, since $\theta_{\rm out}=0$ at $r^2=x_{\rm BH}[2(m_1+m_2)]/(\lambda\tau)$, which is approximately a sphere.  
However, we see that this solution differ from the five-dimensional Reissner-Nordstr\"om-de Sitter solution in the following point; Each $r = {\rm const}$ surface is topologically a lens space $L(2;1)={\rm S}^3/{\mathbb Z}_2$, while in the five-dimensional Reissner-Nordstr\"om-de Sitter solution, it is diffeomorphic to ${\rm S}^3$. We can regard ${\rm S}^3$ and a lens space $L(2;1)={\rm S}^3/{\mathbb Z}_2$  as examples of Hopf bundles i.e. ${\rm S}^1$ bundle over ${\rm S}^2$. The difference between these metrics appears in Eqs.(\ref{eq:near}) and (\ref{eq:lens}): $d\psi$ in ${\rm S}^3$ metric (\ref{eq:near}) is replaced by $d\psi/2$ in $L(2;1)$ metric (\ref{eq:lens}). Therefore, at late time, the topology of the trapped surface is a lens space $L(2;1)={\rm S}^3/{\mathbb Z}_2$.

{}From these results, we see that 
if $m_1+m_2<8/(27\lambda ^2)$ (in this letter, we consider only this case), this solution describes the dynamical situation such that two black black holes with the spatial topologies ${\rm S}^3$ coalesce and convert into a single black hole with the spatial topologies of a lens space $L(2;1)={\rm S}^3/{\mathbb Z}_2$. We should note that in the case of the five-dimensional Kastor-Traschen solution~\cite{KT,London}, in the contracting phase, two black holes with the topology of ${\rm S}^3$ coalesce into a single black hole with the topology of ${\rm S}^3$.

Finally, in order to compare the area of a single black hole formed by the coalescence of two black holes at late time, let us consider the five-dimensional Kastor-Traschen solution~\cite{KT,London} which has the two black holes with the masses $m_1$ and $m_2$ at early time,
\begin{eqnarray}
ds^2&=&-\left(\lambda \tau+\frac{m_1}{|{\bm r}-{\bm r_1}|^2}+\frac{m_2}{|{\bm r}-{\bm r_2}|^2}\right)^{-2}d\tau^2\nonumber\\
    & &+\left(\lambda \tau+\frac{m_1}{|{\bm r}-{\bm r_1}|^2}+\frac{m_2}{|{\bm r}-{\bm r_2}|^2}\right)
(dr^2+r^2d\Omega_{{\rm S}^3}^2),
\end{eqnarray}
where ${\bm r}=(x,y,z,w)$ is the position vector on the four-dimensional Euclid space, and ${\bm r}_1$ and ${\bm r}_2$ denote the position vectors of the two black holes on the four-dimensional Euclid space. 

Since this metric near the black horizon is equal to that of Eq.(\ref{eq:near}) at early time ($\tau\to -\infty$),
each black hole has the same area as that in our solution.
The induced metrics on the cross sections of the apparent horizon with the $\tau = {\rm const}$ at early time become
\begin{eqnarray}
& &ds^2|_{{\rm early}}^{{\rm KT}}=ds^2|_{{\rm early}}^{{\rm EH}}=(x_{\rm BH}[m_i]+m_i)d\Omega^2_{{\rm S}^3},
\end{eqnarray}
where ${\rm KT}$ and ${\rm EH}$ mean the Kastor-Traschen solution and our Eguchi-Hanson based solution. Therefore, using Eq.(\ref{eq:cubic}), the areas ${\cal A}^{\rm KT}_{\rm early}$ and ${\cal A}_{\rm early}^{\rm EH}$ of them at the early time can be computed as follows,
\begin{eqnarray}
{\cal A}^{\rm KT}_{\rm early}={\cal A}^{\rm EH}_{\rm early}=\frac{2}{\lambda }\left( x_{\rm BH}[m_1]+x_{\rm BH}[m_2]\right){\cal A}_{{\rm S}^3},
\end{eqnarray}
where ${\cal A}_{{\rm S}^3}$ denotes the area of a three-dimensional sphere with unit radius. On the other hand, the induced metrics of the cross sections of the apparent horizon with the $\tau = {\rm const}$ at late time ($\tau\to -0$) become
\begin{eqnarray}
ds^2|^{\rm KT}_{{\rm late}}&=&( x_{\rm BH}[m_1+m_2]+m_1+m_2)d\Omega^2_{{\rm S}^3},\\
ds^2|^{\rm EH}_{{\rm late}}&=&( x_{\rm BH}[2(m_1+m_2)]
+2(m_1+m_2))
d\Omega^2_{{\rm S}^3/{\mathbb Z}_2}.
\end{eqnarray}
Hence, using Eq.(\ref{eq:cubic}), the areas ${\cal A}^{\rm KT}_{\rm late}$ and ${\cal A}_{\rm late}^{\rm EH}$ of them at the late time can be computed as follows,
\begin{eqnarray}
{\cal A}^{\rm KT}_{\rm late}&=&\frac{2}{\lambda} x_{\rm BH}[m_1+m_2]{\cal A}_{{\rm S}^3}\\
{\cal A}^{\rm EH}_{\rm late}&=&\frac{2}{\lambda}x_{\rm BH}[2(m_1+m_2)]\frac{{\cal A}_{{\rm S}^3}}{2},\label{eq:area}
\end{eqnarray}
respectively. We should note that ${\cal A}_{{\rm S}^3}/2$ in Eq.(\ref{eq:area}) reflects the fact that the black hole at late time after the coalescence of the two black holes is topologically a lens space $L(2;1)={\rm S}^3/{\mathbb Z}_2$. 
Thus, we see that if each black holes at early time in our solution have the same area with that in the Kastor-Traschen solution, the ratio of the area of the single black hole at late time in our solution to that in the five-dimensional Kastor-Traschen solution is given by
\begin{eqnarray}
\frac{{\cal A}^{\rm EH}_{\rm late}}{{\cal A}^{\rm KT}_{\rm late}}
=\frac{x_{\rm BH}[2(m_1+m_2)]}{2x_{\rm BH}[m_1+m_2]},
\end{eqnarray}
which is larger than one regardless of the values of 
$m_1$, $m_2$, since $x_{\rm BH}[m]$ is the concave downward and increasing function of $m$.
The Fig.\ref{fig:1} shows how the ratio ${\cal A}^{\rm EH}_{\rm late}/{\cal A}^{\rm KT}_{\rm late}$ depends on the initial total masses $m_1+m_2$ of two black holes at early time $\tau\to-\infty$.
\begin{figure}[htbp]
\begin{center}
\includegraphics[width=0.7\linewidth]{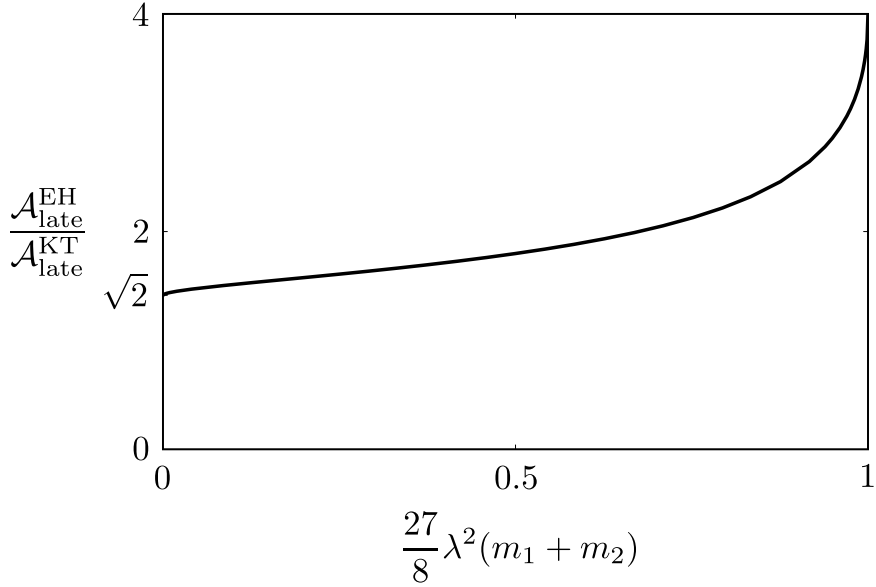}
\end{center}
\caption{This graph shows how the ratio ${\cal A}^{\rm EH}_{\rm late}/{\cal A}^{\rm KT}_{\rm late}$ depends on the total mass parameter $m_1+m_2$.}\label{fig:1}
\end{figure}
 As seen in this figure, the ratio ${\cal A}^{\rm EH}_{\rm late}/{\cal A}^{\rm KT}_{\rm late}$ is monotonically increasing function of the initial total mass of two black holes. We also see that this ratio has the range of $\sqrt{2}< {\cal A}^{\rm EH}_{\rm late}/{\cal A}^{\rm KT}_{\rm late}\le 4$.

There are two main differences between our solution and the Kastor-Traschen solution~\cite{KT,London}:
Firstly, two black holes with the topology of ${\rm S}^3$ coalesce and change into a single black hole with the topology of $L(2;1)={\rm S}^3/{\mathbb Z}_2$, while for the Kastor-Traschen solution, two black holes with the topology with ${\rm S}^3$ coalesce into a single black hole with the topology of ${\rm S}^3$. Secondly, after two black holes coalesce, where we assume that each black hole in our solution has the same mass and area as that in the Kastor-Traschen solution initially, the area of the single black hole formed by the coalescence at late time in our solution is larger than that in the Kastor-Traschen solution. 
 {These differences are essentially due to the asymptotic structure. While the Kastor-Traschen solution is asymptotically de Sitter and each $r = {\rm const}$ surface has the topological structure of ${\rm S}^3$, our solution is asymptotically locally de Sitter and  $R = {\rm const}$ surface is topologically $L(2;1)={\rm S}^3/{\mathbb Z}_2$.} 
 
 In order to know what asymptotic structure our living world admits, it is important to clarify the difference between phenomena which occur in spacetimes with a variety of asymptotic structures. If we can detect the areas after coalescence of higher dimensional two black holes, we would obtain information as to asymptotic structure. Namely, if we find that the total area of two black holes at early time and the area at late time after the coalescence, we can know what the asymptotic structure is.

In this letter, we considered a pair of black holes situated on the north pole 
$\tilde{\theta}=0$ and the south pole $\tilde \theta=\pi$ on ${\rm S}^2$-bolt of 
the Eguchi-Hanson space. Using isometries acting on the ${\rm S}^2$-bolt, we can 
construct multiple black holes on any points of ${\rm S}^2$-bolt. 
It is an interesting future work to see the coalescing process of these 
black holes with the time lapsed as done in Ref.\cite{IN}. 
It is also important to study the coalescence of black holes 
with compact extra dimensions.  
We are successful in constructing a multi-black holes solutions 
on the multi-Taub-NUT space~\cite{IIKMMT} with a positive cosmological 
constant, which would be useful for the study.

We are grateful to K. Matsuno, K. Nakao and Y. Yasui for useful discussions. 
This work is supported by the Grant-in-Aid
for Scientific Research No.14540275 and No.13135208.

\end{document}